\documentclass{elsart3}

\usepackage{epsf}

\usepackage{amssymb}

\begin{document}

\begin{frontmatter}



\title{Randomness Effect on Multicritical
Phenomena in Double-Exchange Systems}


\author[ERATO]{Yukitoshi Motome},
\author[Aogaku]{Nobuo Furukawa},
\author[ERATO,UT,CERC]{Naoto Nagaosa}

\address[ERATO]{
Tokura Spin SuperStructure Project (SSS), ERATO,
Japan Science and Technology Corporation (JST),
c/o National Institute of Advanced Industrial Science and Technology (AIST),
Tsukuba, Ibaraki 305-8562, Japan}
\address[Aogaku]{
Department of Physics, Aoyama Gakuin University, 
Setagaya-ku, Tokyo 157-8572, Japan}
\address[UT]{
Department of Applied Physics, University of Tokyo,
Bunkyo-ku, Tokyo 113-8656, Japan}
\address[CERC]{
Correlated Electron Research Center (CERC), AIST,
Tsukuba, Ibaraki 305-8562, Japan}

\begin{abstract}
Double-exchange model interacting with adiabatic phonons is studied 
in the presence of randomness by using the Monte-Carlo method 
and the systematic size-scaling analysis. 
A bicritical behavior is found between 
the ferromagnetic metal and the charge-ordered insulator. 
We find the distinct response to the randomness  between these two states, 
which agrees well with the experimental results 
in the colossal magnetoresistance manganites.
\end{abstract}

\end{frontmatter}


Phase competition and randomness play a key role 
in the colossal magnetoresistance (CMR) phenomena in manganese oxides
\cite{Tokura2000,Dagotto2001}. 
The former phase competition between the ferromagnetic metal (FM)
and the charge-ordered insulator (CO) leads to large fluctuations 
near the multicritical point
\cite{Tomioka2002}.
The latter randomness suppresses the transition temperatures 
above which CMR is much enhanced
\cite{AkahoshiPREPRINT}.
Theoretically, except for a phenomenological argument 
\cite{Burgy2001}, 
this multicritical phenomenon has not been fully investigated thus far. 
Therefore, it is strongly desired to study a realistic model 
for this phenomenon including the metal-insulator transition. 

In this work, we study the multicritical phenomena 
in the double-exchange (DE) model interacting with phonons 
whose Hamiltonian reads
\begin{eqnarray}
H = &-& t \sum_{\langle ij \rangle} (1 + S_i S_j) 
( c_{i}^\dagger c_{j} + {\rm h.c.} ) 
\nonumber \\
&-& g \sum_i c_i^\dagger c_i \ x_i + \frac12 \sum_i x_i^2 
+ \frac{\lambda}{2} \sum_{\langle ij \rangle} x_i x_j 
\nonumber \\
&+& \sum_i \varepsilon_i \ c_i^\dagger c_i + \gamma \ M^2 X^2. 
\label{eq:H}
\end{eqnarray}
Here, the first line is the DE part which is derived 
in the limit of the large Hund coupling with the Ising symmetry
($S_i = \pm 1$)
\cite{Motome2001}, and
the second line represents the electron-phonon coupling and 
the elastic energy of phonons. 
We consider the classical phonon in the breathing mode. 
The $\lambda$ term is for the cooperative distortion of 
the MnO$_6$ tetrahedra. 
The first term in the last line denotes 
the random on-site potential. 
We consider here the binary-type distribution 
$\varepsilon_i = \pm \Delta$. 
The last term is a Gintzburg-Landau-type term which describes 
the competition between FM of the DE origin and 
CO of the checker-board type, whose order parameters are denoted 
as $M$ and $X$, respectively. 
We consider that this term mimics the elements 
beyond this simple model such as the orbital degeneracy and 
the antiferromagnetic coupling between localized spins. 
We consider model (\ref{eq:H}) at the electron density $n=0.5$
on the square lattice.
We set the halfbandwidth $W=4t=1$ as an energy unit, and 
take $\lambda=0.1$ and $\gamma=1$ in the following.

We study the thermodynamics in model (\ref{eq:H}) by MC method
\cite{Yunoki1998}. 
We have typically 10000 samplings for measurement 
after 1000 steps for thermalization 
in the absence of the randomness. 
In the presence of the randomness, the measurement is typically performed 
1000 times for a configuration of the random potential $\{ \varepsilon_i \}$, 
and the results are random-averaged for typically 16 different configurations. 
The transition temperatures for FM and CO are determined 
by the Binder parameter as well as  the inflection point of 
the temperature dependence of each order parameter. 
Figure~\ref{fig:Binder,inflection} shows a part of these analyses ($\Delta=0$). 
Both results consistently indicate 
the emergence of the long-range ordering. 

\begin{figure}
\epsfxsize=8cm
\centerline{\epsfbox{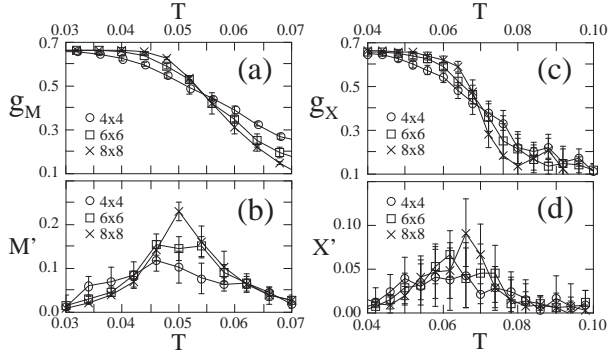}}
\caption{
The Binder parameter and 
the numerical derivatives of the temperature dependence of the order parameter. 
(a) and (b) are for FM at $g=0.8$, and (c) and (d) are for CO at $g=1.2$, 
respectively. 
Critical temperatures are indicated by 
the crossing of the Binder parameter in (a) and (c), and
the peak of the numerical derivatives in (b) and (d).
} 
\label{fig:Binder,inflection} 
\end{figure}

\begin{figure}
\epsfxsize=6.5cm
\centerline{\epsfbox{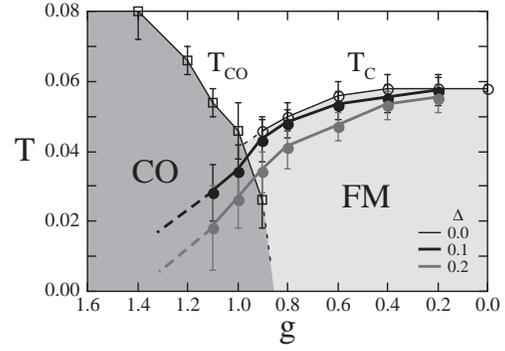}}
\caption{
Phase diagram of model (\ref{eq:H}). 
} 
\label{fig:phase} 
\end{figure}

Figure~\ref{fig:phase} displays the phase diagram of model (\ref{eq:H}).  
In the pure case $\Delta=0$, the phase diagram shows the bicritical behavior 
where FM and CO phases touch with each other at $g \simeq 0.9$. 
When we introduce the randomness, these two phases show a distinct response; 
FM is rather robust whereas CO is surprisingly fragile to the randomness. 
Consequently, the CO phase is destroyed rapidly, and instead the FM state 
extends to larger-$g$ regions. 
This is a consequence of the phase competition near the bicritical point; 
if one phase diminishes, the other takes over. 
These aspects well agrees with the experimental results in the CMR manganites
\cite{AkahoshiPREPRINT}. 
Therefore, we believe that model (\ref{eq:H}) captures 
the essential physics of this multicritical phenomenon in manganites. 
In experiments, the enhanced CMR is observed near the region 
where FM replaces CO
\cite{AkahoshiPREPRINT}. 
Transport property and electronic structure in the present model 
will provide a key to understand the mechanism of CMR. 

\end{document}